\documentclass[5p]{elsarticle}
\usepackage{amsmath,amsfonts}
\usepackage{hyperref}
\usepackage{subfigure}
\usepackage{graphicx}
\usepackage{xspace}

\graphicspath{{figures/}{../../su3/figures/}}

\newcommand{\Eq}[1]{Eq.~(\ref{#1})}

\newcommand{\sigx}{\ensuremath{\sigma_q}}
\newcommand{\sigy}{\ensuremath{\sigma_s}}
\newcommand{\bsig}{\ensuremath{\bar{\sigma}}}
\newcommand{\bvarphi}{\ensuremath{{\boldsymbol\varphi}}}

\newcommand{\tsig}{\ensuremath{\tilde{\sigma}}}
\newcommand{\bsigx}{\ensuremath{\bar{\sigma}_q}}
\newcommand{\bsigy}{\ensuremath{\bar{\sigma}_s}}
\newcommand{\tsigx}{\ensuremath{\tilde{\sigma}_q}}
\newcommand{\tsigy}{\ensuremath{\tilde{\sigma}_s}}
\newcommand{\bOmega}{\ensuremath{\bar{\Omega}}}
\newcommand{\tOmega}{\ensuremath{\tilde{\Omega}}}
\newcommand{\ds}{\displaystyle}

\usepackage{xspace}
\newcommand{\lsm}{L\ensuremath{\sigma}M\xspace}
\newcommand{\adolc}{ADOL-C\xspace}

\newcommand{\einh}[1]{\ensuremath{\,\text{#1}}}
\newcommand{\MeV}{\einh{MeV}}

\newcommand{\R}{{\mathbb R}}

\begin{document}

\title{On the efficient computation of high-order derivatives for
implicitly defined functions}

\author{Mathias Wagner}
\ead{mathias.wagner@physik.tu-darmstadt.de}
\address{Institut f\"{u}r Kernphysik, TU Darmstadt, D-64289
Darmstadt, Germany}
\address{ExtreMe Matter Institute EMMI,
GSI Helmholtzzentrum f\"{u}r Schwerionenforschung GmbH,
D-64291 Darmstadt, Germany}

\author{Andrea Walther}
\ead{Andrea.Walther@uni-paderborn.de}
\address{Institut f\"{u}r Mathematik, Universit\"{a}t Paderborn, D-33098 Paderborn, Germany}  

\author{Bernd-Jochen Schaefer}
\address{Institut f\"{u}r Physik, Karl-Franzens-Universit\"{a}t,
A-8010 Graz, Austria}
\ead{bernd-jochen.schaefer@uni-graz.at}

\begin{abstract}
Scientific studies often require the precise calculation of
derivatives. In many cases an analytical calculation is not
feasible and one resorts to evaluating derivatives numerically.
These are error-prone, especially for higher-order derivatives. A
technique based on algorithmic differentiation is presented which
allows for a precise calculation of higher-order derivatives. The
method can be widely applied even for the case of only numerically
solvable, implicit dependencies which totally hamper a semi-analytical
calculation of the derivatives. As a demonstration the method is
applied to a quantum field theoretical physical model. The
results are compared with standard numerical derivative methods.
\end{abstract}

\maketitle

\section{Physical motivation}

In many scientific studies the knowledge of derivatives of a
given quantity is of particular importance. For example in theoretical
physics, especially in thermodynamics, many quantities of interest
require the calculation of derivatives of an underlying thermodynamic
potential with respect to some external parameters such as
temperature, volume, or chemical potentials. In many cases the
thermodynamic potentials can only be evaluated numerically and one is
forced to employ numerical differentiation techniques which are
error-prone as any numerical methods. Furthermore, the thermodynamic
potential has to be evaluated at the physical point defined by
minimizing the thermodynamic potential with respect to some
condensates yielding the equations of motion (EoM). Generally,
these equations can be solved only numerically and thus introduce
additional implicit dependencies which makes the derivative
calculations even more complicated.

Even in cases where the thermodynamic potential and the implicit
dependencies on the external parameters are known analytically, the
evaluation of higher-order derivatives becomes very complex and
tedious and in the end impedes their explicit calculation.

In this work we present a novel numerical technique, based on
algorithmic differentiation (AD) to evaluate derivatives of arbitrary
order of a given quantity at machine precision. Compared to other
differentiation techniques such as the standard divided
differentiation (DD) method or symbolic differentiation, the AD
produces truncation-error-free derivatives of a function which is
coded in a computer program. Additionally, AD is fast and reduces the work required for analytical calculations and coding, especially for higher-order derivatives. Furthermore,
the AD technique is applicable even if the implicit dependencies on
the external parameters are known only numerically. In
Ref.~\cite{GrWa08} a comprehensive introduction to AD can be found. First remarks about the computation of derivative of implicitly defined
functions were already contained in~\cite{Kedem1980}. However, a detailed
description and analysis is not available yet.
Additional information about tools and literature on AD are available
on the web-page of the AD-community~\cite{AD}.

This work is organized in the following way: For illustrations we will
introduce an effective model, the so-called linear sigma model with
quark degrees of freedom in Sec.~\ref{sec:model}. This model is widely used for the description of the
low-energy sector of strongly interacting matter.
As a starting point the basic thermodynamic grand
potential and the EoM of this model are calculated in a simple
mean-field approximation in order to elucidate the technical problems
common in such types of calculations. Before we demonstrate the power
of the AD method by calculating certain Taylor expansion
coefficients up to very high orders for the first time in
Sec.~\ref{sec:taylor}, the AD method itself and some mathematical
details are introduced in Sec.~\ref{sec:ad}. Details for the
calculation of higher-order derivatives of implicit functions are
given in the following Sec.~\ref{sec:impfun}. In 
Sec.~\ref{sec:ADvsDD} the results of the AD method are confronted with the
ones of the standard divided differences (DD) method in order to
estimate the truncation and round off errors. Finally, we end with a
summary and conclusion in Sec.~\ref{sec:summary}.

\section{A model example}
\label{sec:model}

In order to illustrate the key points of the AD method we employ
a quantum field theoretical model \cite{Gell-Mann:1960np}. This model
can be used to investigate the phase structure of strongly interacting
matter described by the underlying theory of Quantum Chromodynamics (QCD).
Details concerning this effective linear sigma model (\lsm) in the QCD
context can be found in reviews, see e.g.~\cite{MeyerOrtmanns:1996ea,
Schaefer:2006sr}. 

The quantity of interest for the exploration of the phase
structure is the grand potential of the \lsm. This thermodynamic
potential depends on the temperature $T$ and quark chemical potential
$\mu$ because the particle number can also vary. It is calculated in
mean-field approximation whose derivation for three quark flavors is
shown explicitly in \cite{Schaefer:2008hk}. For the \lsm the total
grand potential $\Omega$ consists of two contributions
\begin{equation}
\label{eq:grand_pot}
\Omega(T,\mu;\sigx,\sigy)  = U \left( \sigx,\sigy \right) +
\Omega_{\bar{q}q}(T,\mu;\sigx,\sigy)\ , 
\end{equation}
where the first part, $U$, stands for the purely mesonic potential
contribution and is a function of two condensates, $\sigx$ and
$\sigy$. The second part, $\Omega_{\bar q q}$, is the quark
contribution and depends on the two condensates as well as on the
external parameters temperature $T$ and, for simplicity, only one
quark chemical potential $\mu$. Since the quark contribution arises
from a momentum-loop integration over the quark fields, it is given by
an integral which cannot be evaluated in closed form analytically.
Readers who are unfamiliar with the physical
details, may simply regard \Eq{eq:grand_pot} as an only
numerically known function and continue with the reading above
\Eq{eq:eom}, which introduces an implicit dependency on the
parameters $T$ and $\mu$ whose treatment with the AD technique is the
major focus of this work.

Explicitly, in mean-field approximation the quark contribution reads
\begin{multline}
\label{eq:quark_pot}
\Omega_{\bar{q}q} = 6 T \sum_{f=u,d,s} \int\limits_0^\infty \!
\frac{d^3 k}{(2\pi)^3} \left\{ \ln  \left(1-n_{q,f}(T,\mu)\right) \right.\\
+ \left.\ln
 (1-{\bar n}_{\bar{q},f}(T,\mu)) \right\}\ ,
 \end{multline}
where a summation over three quark flavors $f$ is included. The usual
fermionic occupation numbers for the quarks are denoted by
\begin{equation}
n_{q,f} (T,\mu )=\frac{1}{1+\exp\left((E_{q,f}-\mu)/T\right)}
\end{equation}
and for antiquarks by
${\bar n}_{\bar q,f}(T,\mu) \equiv n_{q,f} (T,-\mu)$ respectively. In
this example only two different single-particle energies,
$E_{q,i},\ i=q,s$, emerge
\begin{equation}
E_{q,q}= \sqrt{k^2 + (g \sigx /2)^2}\quad \text{and} \quad E_{q,s}=
\sqrt{k^2 + (g \sigy /\sqrt{2})^2}\ . 
\end{equation}
The first index $i=q$ denotes the combination of two mass-degenerate
light-quark flavors ($u, d$) and the other index $s$ labels the heavier strange
quark flavor. The expressions in parentheses in $E_{q,i}$ are the
corresponding quark masses. In this way, the dependency of the grand
potential on the condensates, $\sigma_i$, $i=q,s$ enter through the
quark masses, which has not been indicated explicitly in
\Eq{eq:quark_pot}.

The mesonic potential does not depend on the quark chemical potential
nor on the temperature explicitly. It is just a function of the two
condensates and reads
\begin{multline}
\label{eq:umeson}
U(\sigx,\sigy) = \frac{m^{2}}{2}\left(\sigx^{2} +
\sigy^{2}\right) -h_{q} \sigx -h_{s} \sigy
- \frac{c}{2 \sqrt{2}} \sigx^2 \sigy
\\
+ \frac{\lambda_{1}}{2} \sigx^{2} \sigy^{2}+
\frac{1}{8}\left(2 \lambda_{1} +
\lambda_{2}\right)\sigx^{4}+\frac{1}{8}\left(2 \lambda_{1} +
2\lambda_{2}\right) \sigy^{4}\ ,
\end{multline}
wherein all remaining quantities, e.g.~$m, h_q, \ldots$ are constant
parameters.

Since the physical condensates, $\bsigx$ and $\bsigy$, are determined
by the extrema (minima) of the total grand potential with respect to
the corresponding fields, they fulfill the equations of motion
\begin{equation}
\label{eq:eom}
\left. \frac{\partial \Omega(T,\mu;\sigx,\sigy))}{\partial
   \sigma_i}\right|_{\scriptsize \begin{array}{l}\sigx=\bsigx,\\\sigy=\bsigy
 \end{array}} = 0\ ; \qquad i=q,s\ .
\end{equation}

This in turn introduces an implicit $T$- and $\mu$-dependence of both
condensates,
\begin{equation}
\bsig_i = \bsig_i(T,\mu)\ ; \qquad i=q,s\ .
\end{equation}

These quantities represent the physical order parameters which,
together with the grand potential, are the basis of the exploration of the phase
structure of the model. We denote the grand potential evaluated at
$\sigma_i=\bsig_i,$ $i=q,s$, as
\begin{equation}
\bOmega(T,\mu) = \Omega\left(T,\mu; \bsigx(T,\mu), \bsigy(T, \mu)\right)
\end{equation}

In order to find the temperature and chemical potential behavior of
the order parameters the integral in \Eq{eq:quark_pot} and
simultaneously the EoM have to be solved numerically. This already
is an example suitable for an AD application, because a
derivative of a only numerically solvable, implicit function is needed as input.
Later we will be interested in higher-order derivatives of the grand
potential with respect to, e.g., the chemical potential. For example,
the quark number density at the physical point is defined by
\begin{equation}
\rho_q (T,\mu)= - \frac{\partial \bOmega(T,\mu)}{\partial \mu} \ .
\end{equation}
In cases without an implicit $T$- or $\mu$-dependence in the
thermodynamic potential some progress can be made by calculating the
corresponding derivatives explicitly and solving the corresponding
equations numerically. This might be feasible for lower-order
derivatives, in particular, if parts of the derivative calculations
can be performed by some computer algebra packages like Mathematica or
Maple. But for higher-order derivatives this procedure is error-prone
and time-consuming and not applicable anymore.

In the following sections the algorithmic differentiation
technique for implicitly defined functions is introduced on a general
mathematical level.

\section{Algorithmic Differentiation}
\label{sec:ad}

Suppose the function ${\bf F}:\R^n \mapsto \R^m$, ${\bf y} ={\bf F} ({\bf x})$ describing an
arbitrary algebraic mapping from $\R^{n}$ to $\R^m$ is defined by an
evaluation procedure in a high-level computer language like Fortran or
C. The technique of algorithmic differentiation provides derivative
information of arbitrary order for the code segment in the computer that evaluates 
${\bf F}({\bf x})$ within working accuracy.
For this purpose, the basic differentiation rules such as, e.g., the product rule 
are applied to each statement of the given code segment.
This local derivative information is then combined by the chain rule to calculate 
the overall derivatives.
Hence the code is decomposed into a long sequence of simple evaluations, e.g.,~additions,
multiplications, and calls to elementary functions such as $\sin(x)$
or $\exp(x)$, the derivatives of which can be easily calculated. Exploiting the chain rule yields
the derivatives of the whole sequence of statements with respect to
the input variables. 


As an example, consider the function ${\bf F} : \R^3 \rightarrow \R^2$ with
\begin{align*}
y_1 &= \sin (x_1 * x_2) \\
y_2 &= x_3 + \cos (x_1 * x_2)
\end{align*}
that can be evaluated by the pseudo-code given on the left column of 
Tab.~\ref{fig1}.
On the right-hand side, the resulting statements for the derivative calculation
$\dot{\bf y} = {\bf F}'({\bf x}) \dot{\bf x}$ are given.

\begin{table}[hbt!]
\[
\begin{array}{ll}
v_1 = x_1 * x_2 \phantom{XXXX} & \dot{v}_1 = \dot{x}_1 * x_2 + x_1 * \dot{x}_2 \\
v_2 = \sin (v_1) & \dot{v}_2 = \cos (v_1) * \dot{v}_1 \\
v_3 = \cos (v_1) & \dot{v}_3 = - \sin (v_1) * \dot{v}_1 \\
v_4 = x_3 + v_3 & \dot{v}_4 = \dot{x}_3 + \dot{v}_3 \\
y_1 = v_2 & \dot{y}_1 = \dot{v}_2 \\
y_2 = v_4 & \dot{y}_2 = \dot{v}_4
\end{array}
\]
\caption{Function and derivative calculation.}
\label{fig1}
\end{table}
For the vector $\dot{\bf x} = (1, 0, 0)^T$ one obtains
the first column of the Jacobian $\nabla {\bf F}({\bf x})$ the vector $\dot{\bf y} = (\cos (x_1 * x_2) * x_2$,
$- \sin (x_1 * x_2) * x_2)^T$.
Correspondingly, the other unit vectors in $\R^3$ yield the other two remaining
columns of the Jacobian ${\bf F}'({\bf x})$.

Table~\ref{fig1} illustrates the so-called forward mode of AD, where the derivatives are 
propagated together with the function evaluation.
Alternatively, one may propagate the derivative information from the
dependents ${\bf y}$ to the independents ${\bf x}$ yielding the
so-called reverse mode of AD.


Over the past decades, extensive research activities led to a
thorough understanding and analysis of these two basic modes of AD, where
the complexity results with respect to the required runtime are based on the 
operation count $O_{\bf F}$,~i.e.,~the number of floating point operations required to evaluate
${\bf F} ({\bf x})$, and the degree $d$ of the computed derivatives.
Using the forward mode, one computes the required derivatives together
with the function evaluation in one sweep as illustrated above. The forward mode yields one
{\em column} of the Jacobian $\nabla {\bf F}$ at no more than three
times $O_{\bf F}$ ~\cite{GrWa08}. One {\em row} of $\nabla {\bf F}$,
e.g., the gradient of a scalar-valued component function of ${\bf F}$,
is obtained using the reverse mode in its basic form also at no more
than four times $O_{\bf F}$ \cite{GrWa08}. It is important to note
that this bound for the reverse mode is completely independent of the
number $n$ of input variables. This observation is called 
{\em cheap gradient result}.

For the application discussed in the present work, the forward
mode has been chosen for the efficient computation of higher-order
derivatives which is illustrated in the following paragraphs. To this end, we
consider Taylor polynomials of the form

\begin{align}
\label{eq:x} {\bf x}(t) \equiv &\sum_{j=0}^d{\bf x}_j\, t^j \; : \; \R
\; \mapsto \; \R^n \nonumber \\
&\mbox{ where } \\
\qquad {\bf x}_j  &= 
\frac{1}{j!} \left . \frac{\partial ^j}{\partial t^j} {\bf x}(t)\; \right
|_{t=0}\nonumber
\end{align} 
are scaled derivatives at $t=0$. The expansion is truncated at the
highest derivative degree $d$ which is chosen by the user. The vector
polynomial ${\bf x}(t)$ describes a path in $\R^n$ which is
parameterized by $t$. Thus, the first two vectors $\bf x_1$ and
$\bf x_2$ represent the tangent and the curvature at the base point
${\bf x}_0 = {\bf x}(0)$. Assuming that the function 
${\bf y} = {\bf F} ({\bf x})$
is sufficiently smooth, i.e., $d$ times continuously differentiable, one
obtains a corresponding value path
\begin{align} 
\label{eq:ytay}
{\bf  y}(t) & \equiv \sum_{j=0}^d {\bf  y}_j\, t^j = {\bf  F}({\bf x}(t)) + O(t^{d+1}) \in \R^m.
\end{align}  
The coefficient functions ${\bf y}_j$ are uniquely and smoothly
determined by the coefficient vectors ${\bf x}_i$ with $i\le j$. To
compute this higher-order information, first we will examine for a
given Taylor polynomial
\begin{align*}
{\bf  x}(t) & = {\bf   x}_0 + {\bf  x}_1\,t + {\bf  x}_2\, t^2+ \dots
+ {\bf  x}_{d}\,t^{d}
\in \R^n
\end{align*}
the derivative computation based on ``symbolic'' differentiation. 

Let us generalize the previous relation
${\bf y}(t) = {\bf F}({\bf x}(t))$ given in  Eq.~(\ref{eq:ytay}), and consider
now a general smooth function ${\bf v}(t) = \bvarphi({\bf x}(t))$ as
for example the evaluation of a $\sin(.)$-function. This function
${\bf v}(t)$ represents one of the intermediate values computed during
the function evaluation as illustrated in Table~\ref{fig1}. One obtains
for the Taylor coefficients
\begin{align*}
{\bf  v}_j = \frac{1}{j!}\left.\frac{\partial ^j}{\partial t^j} {\bf v}(t)\;\right|_{t=0}
\qquad \mbox{ and }\qquad
{\bf  \bvarphi}_j(.) = \frac{1}{j!}\bvarphi^{(j)}(.)
\end{align*}
the derivative expressions
\begin{align*}
{\bf  v}_{0} & =  \bvarphi({\bf x}_{0})\\
{\bf  v}_{1} & =  \bvarphi{_1}({\bf x}_{0}) \, {\bf x}_{1}\\
{\bf  v}_{2} & =  \bvarphi{_2}({\bf x}_{0}) \, {\bf x}_{1} \, {\bf x}_{1} + 
              \bvarphi{_1}({\bf x}_{0}) \, {\bf x}_{2}\\ 
{\bf  v}_{3} & =  \bvarphi{_3}({\bf x}_{0}) \, {\bf x}_{1} \, {\bf x}_{1} \, {\bf x}_{1} + 2\,
       \bvarphi{_2}({\bf x}_{0}) \, {\bf x}_{1} \, {\bf x}_{2} + 
       \bvarphi{_1}({\bf x}_{0}) \, {\bf x}_{3} \\ 
{\bf  v}_{4} & =  \bvarphi{_4}({\bf x}_{0}) \, {\bf x}_{1} \, {\bf x}_{1} \, {\bf x}_{1} \, {\bf x}_{1} + 
       3\bvarphi{_3} ({\bf x}_{0}) \, {\bf x}_{1} \, {\bf x}_{1} \, {\bf x}_{2}\\ 
 & \phantom{=}   +\bvarphi{_2}({\bf x}_{0}) \, ({\bf x}_{2} \, {\bf x}_{2}
+ 2\,{\bf x}_{1} \, {\bf x}_{3}) + \,\bvarphi{_1}({\bf x}_{0}) \, {\bf x}_{4} \\
& \phantom{=} \vdots 
\end{align*}
Hence, the overall complexity grows rapidly with the degree $d$ of the Taylor 
polynomial. To avoid these prohibitively expensive calculations the standard
higher-order forward sweep of algorithmic differentiation is based on  Taylor arithmetic
\cite{BK78} yielding an effort that grows like $d^2$ times the 
cost of evaluating $\bvarphi({\bf x})$. This is quite obvious for arithmetic
operations such as multiplications or additions, where one obtains
the recursion shown in Table~\ref{tab:arith},
\begin{table}
\begin{center}
\begin{tabular}{c|l|c|c}
$v(t) =$ & $\;$Formula $(1\le k\le d)$ &OPS & MOVES\\ 
\hline 
$x(t) + y(t)\;$ & $\quad v_{k} = x_{k}+y_{k}$                         & $\sim 2d$ & $3d$ \\ \hline  
$x(t) * y(t)\;$ & $\quad v_{k} = \sum\limits_{j=0}^{k} x_{j}*y_{k-j}$  & $\sim d^2$ & $3d$ \\ \hline  
\end{tabular}
\end{center}
\caption{Taylor coefficient propagation for arithmetic operations}
\label{tab:arith}
\end{table}
where OPS denotes the total number of floating point operations and
MOVES the total number of memory accesses required to compute all
Taylor coefficients $v_0,\ldots,v_d$. For a general elemental function
$\varphi$, one finds also a recursion with quadratic complexity by
interpreting $\varphi$ as solution of a linear ordinary differential equation
as described in
\cite{GrWa08}. Table~\ref{tab:uni} illustrates the resulting
computation of the Taylor coefficients for the exponential function
\begin{table}
\begin{center}
\begin{tabular}{c|l|c|c}
$v(t) =$ & $\;$Formula $(1\le k\le d)$ & OPS & MOVES\\ 
\hline 
$\exp(x(t))$ &  $\quad k v_{k} = \sum\limits_{j=1}^{k}  j v_{k-j}  x_{j}$ & $\sim d^2$  & $2 d$ 
\end{tabular}
\end{center}
\caption{Taylor coefficient propagation for exponential}
\label{tab:uni}
\end{table}
Similar formulas can be found for all intrinsic functions. This fact
permits the computation of higher-order derivatives for the vector
function ${\bf F}({\bf x})$ as composition of elementary components.
\medskip

\noindent
The AD-tool ADOL-C \cite{ADOL-C} uses the Taylor arithmetic as
described above to provide an efficient calculation of higher-order
derivatives.
\section{Higher-order Derivatives of Implicit Functions}
\label{sec:impfun}
\subsection{Basic Algorithm}
\label{sub:basic}

\noindent
For the application considered here, higher-order derivatives of a
variable ${\bf y } \in \R^{m}$ are required, where ${\bf y}$ is 
{\em implicitly} defined as a
function of some variable ${\bf x} \in \R^{n-m}$ by an algebraic
system of equations
\begin{align*}
{\bf G}({\bf z}) \; = \; 0 \in \R^m \quad
{\rm with} \quad {\bf z} = ({\bf y}, {\bf x}) \in \R^n .
\end{align*}

Naturally, the $n$ arguments of ${\bf G}$ need not be partitioned in
this regular fashion. To provide flexibility for a convenient
selection of the $p \equiv n-m$ {\em truly} independent variables
$\bf x$, let ${\bf P} \in \R^{p\times n}$ be a projection matrix with
only $0$ or $1$ entries that picks out these independent variables. 
Hence, ${\bf P}$ is a column permutation of the matrix
$[0,{\bf I}_{p}] \in \R^{p\times n}$. Then the nonlinear system
\begin{align*}
{\bf G}({\bf z}) \; = \; 0, \quad {\bf P z} =  {\bf x},
\end{align*} 
has a regular Jacobian, wherever the implicit function theorem yields
${\bf y}$ as a function of ${\bf x}$. Therefore, we may also write with
${\bf H}:\R^n \mapsto \R^n$
\begin{align}
\label{eq:F}
{\bf H}({\bf z}) \equiv \left(\begin{array}{c}
                   {\bf G}({\bf z}) \\
                   {\bf P} {\bf z}
                 \end{array} \right)\; = \;
           \left(\begin{array}{c}
                   0 \\
                   {\bf P z}
                 \end{array} \right)\; = \; {\bf S\, x},
\end{align}
for the seed matrix ${\bf S} = [0,{\bf I}_p]^\top \in \R^{n \times p}$. 
Now, we have rewritten the original implicit functional relation between 
${\bf x}$ and ${\bf z}$ as an inverse relation ${\bf H}({\bf z}) = {\bf S x}$.
Assuming an ${\bf H} : \R^n \mapsto \R^n $ that
is locally invertible we can evaluate the required derivatives
of the implicitly defined ${\bf z} \in \R^n$ with respect to
${\bf x} \in \R^p$ using the computation of higher-order derivatives
described above in the following way.


Starting with a Taylor expansion Eq. (\ref{eq:x}) of ${\bf x}$ 
and a corresponding
solution ${\bf z} ({\bf x}(t))$ of Eq. (\ref{eq:F}), one obtains 
for a sufficiently smooth ${\bf H}$ the representation
\[
{\bf Sx} = {\bf H}({\bf z}({\bf x}(t))) = \sum^d_{j=0} {\bf H}_j t^j + O (t^{d+1}).
\]
Substituting the Taylor expansion of ${\bf x}$ into the previous equation yields
\[
{\bf S} \sum^d_{j=0} {\bf x}_j t^j = \sum^d_{j=0} {\bf H}_j t^j + O(t^{d+1}).
\]%
From the comparison of coefficients, it follows that 
\begin{equation}
\label{eq:2}
{\bf Sx}_{j} t^j = {\bf H}_j t^j\; \Leftrightarrow \; {\bf Sx}_j = {\bf H}_j = \frac{1}{j!}
\left(\frac{\partial^j}{\partial t^j} {\bf H}({\bf z}({\bf x}(t)))\right)\bigg|_{t=0}.
\end{equation}
As a next step, the structure of the Taylor coefficients ${\bf H}_j$ is analyzed.
For the first three coefficients, one has 
\begin{align*}
{\bf H}_0 &= {\bf H}({\bf z}({\bf x}(0))) = {\bf H}({\bf z}_0) \\
{\bf H}_1 &= \left(\frac{\partial}{\partial t} {\bf H} ({\bf z}({\bf x}(t)))\right)\bigg|_{t=0} = {\bf H_z} ({\bf z}_0) {\bf z}_1 \\
{\bf H}_2 &= \frac{1}{2} \left(\frac{\partial^2}{\partial t^2} {\bf H}({\bf z}({\bf x}(t)))\right)\bigg|_{t=0} \\
& = \frac{1}{2} \left(\frac{\partial}{\partial t} 
{\bf H_z}({\bf z}({\bf x}(t))) \frac{\partial}{\partial t} {\bf z}({\bf x}(t))\right)\bigg|_{t=0} \\
&= {\bf H_z} ({\bf z}_0) {\bf z}_2 + \frac{1}{2} {\bf H_{zz}} ({\bf z}_0) {\bf z}_1 
{\bf z}_1
\end{align*}
due to the definition of ${\bf z}_j$,
where ${\bf H_z}(.)$ denotes the derivative of ${\bf
H}(.)$ with respect to its argument.
For the higher-order coefficients, it is now shown that they have the structure
\begin{equation}
\label{eq:14}
{\bf H}_j = \frac{1}{j!} \left( {\bf H_z}({\bf z}({\bf x}(t))) \frac{\partial^j}{\partial t^j} {\bf z}(t)\right)\bigg|_{t=0}
+ {\bf \tilde{H}}_j ({\bf z}({\bf x}(t)))|_{t=0}
\end{equation}
where ${\bf \tilde{H}}_j({\bf z}({\bf x}(t)))$ involves only derivatives of order $j-1$ with 
respect to $t$ and hence
\[
{\bf \tilde{H}}_j({\bf z}({\bf x}(t)))|_{t=0} = {\bf \hat{H}}_j ({\bf z}_0, \ldots , 
{\bf z}_{j-1}).
\]

\noindent
For ${\bf H}_2$, one obtains
\[
{\bf \tilde{H}}_2 ({\bf z}({\bf x}(t))) = \frac{1}{2} 
\left( \frac{\partial}{\partial t} 
{\bf H_z} ({\bf z}({\bf x}(t)))\right) \left( \frac{\partial}{\partial t} {\bf z}({\bf x}(t)) \right).
\]
Now, let the assumption hold for $j-1$.
Then, one obtains for $j$ the equation
\begin{align*}
{\bf H}_j & = \frac{1}{j!}\left(  \frac{\partial^j}{\partial t^j} {\bf H}
({\bf z}({\bf x}(t))) \right)\bigg|_{t=0}\\
   & = \frac{1}{j!} \left(\frac{\partial}{\partial t}
 \frac{\partial^{j-1}}{\partial t^{j-1}} {\bf H} ({\bf z}({\bf x}(t))) \right) \bigg|_{t=0} \\
& = \frac{(j-1)!}{j!} \left[\frac{\partial}{\partial t} 
\left( {\bf H_z} ({\bf z}({\bf x}(t))) \frac{\partial^{j-1}}{\partial
   t^{j-1}} {\bf z}({\bf x}(t))\right.\right.\\
& \phantom{=.}+\left. \left. {\bf \tilde{H}}_{j-1} ({\bf z}({\bf x}(t))) \right) \right] \bigg|_{t=0}  \\
& = \frac{1}{j} \left({\bf H_z} ({\bf z}({\bf x}(t)))
\frac{\partial^j}{\partial t^j} {\bf z}({\bf x}(t))\right) \bigg|_{t=0}  \\
&\phantom{=.} +\frac{1}{j}
\left[\left( \frac{\partial}{\partial t} {\bf H_z} ({\bf z}({\bf x}(t))) \right)
\left( \frac{\partial^{j-1}}{\partial t^{j-1}} {\bf z}({\bf x}(t)) \right)\right] \bigg|_{t=0} \\
& \phantom{=.}+ \frac{1}{j}\left( \frac{\partial}{ \partial t} {\bf \tilde{H}}_{j-1} ({\bf z}({\bf x}(t)))\right) \bigg|_{t=0}.
\end{align*}
Due to the assumptions, the function
\begin{align*}
{\bf \tilde{H}}_j ({\bf z}({\bf x}(t))) & = \frac{1}{j} 
\left( \frac{\partial}{\partial t} {\bf H_z} ({\bf z}({\bf x}(t))) \right)
\left( \frac{\partial^{j-1}}{\partial t^{j-1}} {\bf z}({\bf x}(t))
\right)\\
&\phantom{=.}+ \frac{1}{j!} \frac{\partial}{\partial t} {\bf \tilde{H}}_{j-1} ({\bf z}({\bf x}(t)))
\end{align*}
involves only derivatives of order $j-1$ with respect to $t$
since ${\bf \tilde{H}}_{j-1} ({\bf z}({\bf x}(t)))$ does only contain derivatives of
order $j-2$ with respect to $t$.
Therefore, (\ref{eq:14}) is proven and it follows that
\begin{equation}
\label{eq:15}
{\bf H}_j = {\bf H_z}({\bf z}_0) {\bf z}_j + {\bf \hat{H}}_j ({\bf z}_0, \ldots , {\bf z}_{j-1})
\end{equation}
due to the definition of ${\bf z}_j$.
Combining (\ref{eq:15}) with (\ref{eq:2}), one obtains the equations
\[
{\bf Sx}_j = {\bf H_z} ({\bf z}_0) {\bf z}_j + {\bf \hat{H}}_j ({\bf
 z}_0, \ldots , {\bf z}_{j-1}) \quad 1 \le j \le  d
\]
and therefore
\[
{\bf z}_j = ({\bf H_z}({\bf z}_0))^{-1} ({\bf Sx}_j - {\bf \hat{H}}_j ({\bf z}_0, \ldots , {\bf z}_{j-1})) \quad 1 \le j \le  d
\]

\noindent
where the Jacobian ${\bf H}_{\bf z}({\bf z}({\bf x}(t)))$ and its
factorization can be reused as long as the argument ${\bf z}({\bf x}(t))$
is the same.
For this purpose, the Jacobian ${\bf H}_{\bf z}({\bf z}({\bf x}(t)))$ can be evaluated
exactly by using the forward mode of AD.

Therefore, it remains to provide the missing 
contributions ${\bf \hat{H}}_j ({\bf z}_0, \ldots , {\bf z}_{j-1})$ to compute
the desired Taylor coefficients ${\bf z}_j$.
One starts with the Taylor expansion
\[
{\bf z}_0 = {\bf z}({\bf x}(0)), \quad {\bf z}_1 = ({\bf H_z} ({\bf z}_0))^{-1} {\bf x}_1, \quad {\bf z}_j = 0 \quad
1 \le j \le  d.
\]
For $j=2, \ldots , d$, one performs the following steps
\begin{enumerate}
\item 
A forward mode evaluation of degree $j$.
Since ${\bf z}_j=0$ this yields only the contribution
${\bf \hat{H}}_j ({\bf z}_0, \ldots , {\bf z}_{j-1})$.
\item
One system solve
\[
{\bf z}_j = ({\bf H_z} ({\bf z}_0))^{-1} ({\bf Sx}_j - {\bf \hat{H}}_j 
({\bf z}_0, \ldots , {\bf z}_{j-1}))
\]
to compute ${\bf z}_j$.
\end{enumerate}

This approach provides the complete set of Taylor
coefficients of the Taylor polynomial ${\bf z}={\bf z}({\bf x})$ that
is defined by a given Taylor polynomial \eqref{eq:x} for ${\bf x}$.
These Taylor coefficients of ${\bf z}={\bf z}({\bf x})$ are computed
for a considerably small number of Taylor polynomials ${\bf x}(t)$ to
construct the desired full derivative tensor for the implicitly
defined function ${\bf z}$ according to the algorithm proposed in
\cite{GrUtWa00}.

\medskip
\noindent
\subsection{A Simple Example}
\label{sub:4.2}
\medskip
\noindent
Consider the following two nonlinear expressions
\begin{align*}
G_1 (z_1, z_2, z_3, z_4) & = z^2_1 + z^2_2 - z^2_3\\
G_2 (z_1, z_2, z_3, z_4) & = \cos \, (z_4) - z_1/z_3 
\end{align*}
describing the relation between the Cartesian coordinates $(z_1, z_2)$
and the polar coordinates $(z_3, z_4)$ in the plane.
Assume, one is interested in the derivatives of the second Cartesian and
the second polar coordinate with respect to the first Cartesian and the
first polar coordinate.
Then one has $n=4$, $m=2$, $p=2$, ${\bf x} =(z_1, z_3)$, and 
${\bf y} = (z_2, z_4)$.
The corresponding projection and seed matrix are
\[
P = \left( \begin{array}{cccc}
1 & 0 & 0 & 0 \\
0 & 0 & 1 & 0
\end{array} \right) \quad \mbox{ and } \quad
S^T = \left( \begin{array}{cccc}
0 & 0 & 1 & 0 \\
0 & 0 & 0 & 1
\end{array} \right).
\]
Provided the argument ${\bf z}$ is consistent in that its Cartesian and polar components 
describe the same point in the plane, one has ${\bf G}({\bf z}) = 0$.
In this simple case, one can derive for the implicitly defined functions
$y_1 = z_2 (z_1, z_3)$ and $y_2 = z_4 (z_1, z_3)$ the desired derivatives
explicitly by symbolic manipulation:
\[
y_1  = \sqrt{z^2_3 - z^2_1} \quad \mbox{ and } \quad
y_2  = \arccos \, (z_1/z_3).
\]


\noindent
The derivatives up to order 3 of $y_1$ will be used to verify the results 
from the differentiation of the implicitly defined functions.
These derivatives have the following representation:
\begin{align}
\label{eq:17}
\frac{\partial y_1}{\partial z_1} &= \frac{- z_1}{\sqrt{z^2_3 -
  z^2_1}}   \nonumber \\ 
\frac{\partial y_1}{\partial z_3}& = \frac{z_3}{\sqrt{z^2_3 - z^2_1}} \nonumber \\
\frac{\partial^2 y_1}{\partial z^2_1} &= \frac{-z^2_1}{(z^2_3 - z^2_1)^{\frac{3}{2}}}
- \frac{1}{\sqrt{z^2_3 - z^2_1}} \nonumber \\
\frac{\partial^2 y_1}{\partial z_1 \partial z_3} &= \frac{z_1 z_3}{(z^2_3 - z^2_1)^{\frac{3}{2}}} \nonumber \\
\frac{\partial^2 y_1}{\partial z^2_3} &=
\frac{- z^2_3}{(z^2_3 - z^2_1)^{\frac{3}{2}}} + \frac{1}{\sqrt{z^2_3 - z^2_1}} \\
\frac{\partial^3 y_1}{\partial z^3_1} &= \frac{-3z^3_1}{(z^2_3 - z^2_1)^{\frac{5}{2}}} - \frac{3z_1}{(z^2_3 - z^2_1)^{\frac{3}{2}}}  \nonumber \\
\frac{\partial^3 y_1}{\partial z^2_1 \partial z_3} &=
\frac{3z^2_1 z_3}{(z^2_3 - z^2_1)^{\frac{5}{2}}} +
\frac{z_3}{(z^2_3 - z^2_1)^{\frac{3}{2}}} \nonumber \\
\frac{\partial^3 y_1}{\partial z_1 \partial z^2_3} &=
\frac{z_1}{(z^2_3 - z^2_1)^{\frac{3}{2}}} -
\frac{3z_1 z^2_3}{(z^2_3 - z^2_1)^{\frac{5}{2}}} \nonumber\\
\frac{\partial^3 y_1}{\partial z^3_3} &=
\frac{- 3z_3}{(z^2_3 - z^2_1)^{\frac{3}{2}}} +
\frac{3z^3_3}{(z^2_3 - z^2_1)^{\frac{5}{2}}} \nonumber
\end{align}

As shown in \cite{GrUtWa00}, these derivatives can be computed efficiently and exactly 
from a considerable small number of univariate Taylor expansions like the
one in Eq. \eqref{eq:ytay}. Furthermore, this article also proposes a specific choice
of the employed Taylor polynomials. For the example considered here,
i.e., $m=2$ and $d=3$, one obtains the Taylor expansions
\[
\begin{array}{l}
{\bf x}^1 = \left( \begin{array}{c}
4 \\ 5
\end{array} \right) +
\left( \begin{array}{c}
3 \\ 0
\end{array} \right) t, \quad 
{\bf x}^2 = \left( \begin{array}{c}
4 \\ 5
\end{array} \right) +
\left( \begin{array}{c}
2 \\ 1
\end{array} \right) t, \\
{\bf x}^3 = \left( \begin{array}{c}
4 \\ 5
\end{array} \right) +
\left( \begin{array}{c}
1 \\ 2
\end{array} \right) t, \quad
{\bf x}^4 = \left( \begin{array}{c}
4 \\ 5
\end{array} \right) +
\left( \begin{array}{c}
0 \\ 3
\end{array} \right) t
\end{array}.
\]
Then, the procedure described in the previous section yields the following Taylor expansions of
${\bf z}({\bf x})$ with the base point ${\bf z}_0 = (4,3,5,0.6435)^T$
\begin{align*}
\begin{array}{l}
{\bf z}^1 = {\bf z}_0 + \left( \begin{array}{r}
3 \\ -4 \\ 0 \\ -1
\end{array} \right) t + 
\left( \begin{array}{c}
\;\; 0 \\ 
\!- \frac{25}{6} \\
\;\; 0 \\
\!- \frac{2}{3}
\end{array} \right) t^2 +
\left( \begin{array}{c}
\;\;0 \\
\!- \frac{50}{9} \\
\;\;0 \\
\;\frac{19}{18}
\end{array} \right) t^3 \\[1cm]
{\bf z}^2 = {\bf z}_0 +
\left( \begin{array}{r}
\;2 \\
-1 \\
\;1 \\
\!- \frac{2}{5}
\end{array} \right) t + 
\left( \begin{array}{c}
\;\;0 \\
\!- \frac{2}{3} \\
\;\;0 \\
\!- \frac{2}{75}
\end{array} \right) t^2 + 
\left( \begin{array}{c}
\;\;0 \\
\!- \frac{2}{9} \\
\;\;0 \\
\!- \frac{46}{1125}
\end{array} \right) t^3 
\end{array}
\end{align*}
\begin{align*}
\begin{array}{l}
{\bf z}^3 = {\bf z}_0 +
\left( \begin{array}{l}
1\\
2\\
2\\
\frac{1}{5}
\end{array} \right) t +
\left( \begin{array}{c}
\;\;0 \\
\!- \frac{1}{6} \\
\;\;0 \\
\!- \frac{8}{75}
\end{array} \right) t^2 +
\left( \begin{array}{c}
\:0 \\
\,\frac{1}{9} \\
\:0 \\
\frac{139}{2250}
\end{array} \right) t^3 \\[1cm]
{\bf z}^4 = {\bf z}_0 +
\left( \begin{array}{r}
0\\
5\\
3\\
\frac{4}{5}
\end{array} \right) t + 
\left( \begin{array}{c}
\;\;0 \\
\!- \frac{8}{3} \\
\;\;0 \\
\!- \frac{68}{73}
\end{array} \right) t^2 +
\left( \begin{array}{c}
0 \\
\frac{40}{9} \\
0 \\
\frac{1508}{1125} 
\end{array} \right) t^3.
\end{array}
\end{align*}
From these numerical values, one can derive the desired derivatives in 
Eq.~(\ref{eq:17}) as given below
\begin{align*}
\ds\frac{\partial y_1}{\partial z_1} &= - \ds \frac{4}{3} = \frac{1}{3} {\bf z}^1_{12} =
\frac{1}{3} * (-4)  \\ 
\ds\frac{\partial y_1}{\partial z_3} &= \frac{5}{3} = \frac{1}{3} {\bf z}^4_{12} = 
\frac{1}{3} * 5 \\[.3cm]
\ds\frac{\partial^2 y_1}{\partial z^2_1} &= - \ds\frac{25}{27} = \frac{2}{9} {\bf z}^1_{22} =
\ds\frac{2}{9} * \left(- \frac{25}{6}\right)\\
\ds\frac{\partial^2 y_1}{\partial z_1 \partial z_3} &= \ds\frac{20}{27} = - \frac{5}{36}
{\bf z}^1_{22} + \frac{1}{4} {\bf z}^2_{22} + \frac{1}{4} {\bf z}^3_{22} - \frac{5}{36} {\bf z}^4_{22} \\[.3cm]\ds\frac{\partial^2 y_1}{\partial z^2_3} &= - \frac{16}{27} = \frac{2}{9} {\bf z}^4_{22} =
\frac{2}{9} *  \left(- \frac{8}{3}\right) \\[.3cm]
\ds\frac{\partial^3 y_1}{\partial z^3_1} &= - \ds\frac{100}{81} = \frac{2}{9} {\bf z}^1_{32} =
\frac{2}{9} * - \frac{50}{9} \\[.3cm]
\ds\frac{\partial^3 y_1}{\partial z^2_1 \partial z_3} &= \ds\frac{95}{81} = - \frac{5}{27}
{\bf z}^1_{32} + \frac{2}{3} {\bf z}^2_{32} - \frac{1}{3} {\bf z}^3_{32} + \frac{2}{27} {\bf z}^4_{32}  \\[.3cm]
\ds\frac{\partial^3 y_1}{\partial z_1 \partial z^2_3} &= - \ds\frac{88}{81} = \frac{2}{27}
{\bf z}^1_{32} - \frac{1}{3} {\bf z}^2_{32} + \frac{2}{3} {\bf z}^3_{32} - \frac{5}{27} {\bf z}^4_{32}\\ 
\ds\frac{\partial^3 y_1}{\partial z^3_3} &= \frac{80}{81} = \frac{2}{9} {\bf z}^4_{32} =
\frac{2}{9}* \frac{40}{9}
\end{align*}
where ${\bf z}^j_{k,l}$ denotes the $l$th component of the $k$th Taylor coefficient
of the Taylor expansion $j$.
As can be seen, the required 24 entries of the first three
derivative tensors can be obtained from four
univariate Taylor expansions.
This computation of tensor entries from the Taylor expansions,
i.e. the exact coefficients for the Taylor coefficients, is derived
and analyzed in detail in \cite{GrUtWa00}.

\subsection{Remarks on Efficiency}
\label{sub:4.3}
\noindent
In the procedure described above, the higher-order forward mode of AD
is applied for each value of $j$ for $j=2, \ldots , d$.
Employing in addition to the higher-order forward mode the higher-order
reverse mode, the number of forward and reverse sweeps can be reduced to 
$\log_2 (d)$.
In this case, the values of the required ${\bf \hat{H}}_j ({\bf z}_0,
\ldots , {\bf z}_{j-1})$
is reconstructed from the information available due to the reverse mode
differentiation.
The AD-tool \adolc provides a corresponding efficient implementation of this
algorithm and will be used in the numerical tests below, where the $\log_2$-
behavior of the higher-order derivative calculation can be observed in the 
measured runtimes.


\section{Applications}
\label{sec:ADvsDD}

In the following the previous general mathematical description of the AD
technique is applied to the model example introduced in the beginning.
Furthermore, the AD results are then compared to those obtained with
the standard divided differentiation (DD) method.

\subsection{Algorithmic Differentiation applied to the model}

The first step for the calculation of the $k$-th order derivatives of the
grand potential with respect to $\mu$,
\begin{equation}
\frac{d^{k}}{d \mu^k} \bOmega \left(T,\mu\right) = \frac{d^{k}}{d \mu^k} \Omega \left(T,\mu,\bsigx(T,\mu),\bsigy(T,\mu)\right)\ , 
\end{equation}
by means of the AD technique requires a suitable formulation of
$\bOmega(T,\mu)$. This can be accomplished by a Taylor
expansion of the condensates $\bsig_{q,s} (T, \mu)$. The required
coefficients, i.e., the derivatives
\begin{equation}
\frac{d^{k}}{d \mu^k}\bsig_{i}(T,\mu)\quad;\quad i=q,s
\end{equation}
can be calculated by applying the technique described in the previous
section for implicit functions. The next step consists in the
calculation of the derivatives of $\bOmega(T,\mu)$ w.r.t.~$\mu$ by
using the Taylor expansions of the condensates. In the following the
procedure will be exemplified in detail.

In this example only one, i.e. $p=1$, truly independent variable
$x = \mu$ is considered and the temperature $T$ plays the role of
a constant parameter. The generalization to mixed derivatives with
respect to $T$ and $\mu$ can also be realized but is omitted for
simplicity.

Firstly, the Taylor coefficients for the condensates $\bsig_{i}$ are
needed. This is done via the inverse Taylor expansion capabilities of
\adolc. For that purpose the following function
\begin{equation} 
{\bf G}({\bf z}) = \begin{pmatrix}
\left. \frac{\partial \Omega(T,\mu; \sigx,\sigy)}{\partial
  \sigx}\right|_{\scriptsize
\begin{array}{l}\sigx=\bsigx,\\\sigy=\bsigy
\end{array}}\\
\left.\frac{\partial \Omega(T,\mu; \sigx,\sigy)}{\partial
  \sigy}\right|_{\scriptsize
\begin{array}{l}\sigx=\bsigx,\\\sigy=\bsigy
 \end{array}}
\end{pmatrix}
\end{equation}
is introduced. The $n=3$ dimensional argument ${\bf z} =({\bf y}(x),x)$ splits into
$m=2$ implicitly defined functions ${\bf y} = (\bsigx,\bsigy)$ and $p=n-m=1$
truly independent variable $x =\mu$, cf.~\Eq{eq:F}. Furthermore, the
projection matrix reads $P =(0,0,1) \in \R^{p \times n}$.

In order to obtain the $\mu$-derivatives of the functions $\bsig_i$
for fixed values of $(T,\mu)=(T_0,\mu_0)$ the following steps are
required:
\begin{enumerate}
\item The numerical solution of the EoM, see \Eq{eq:eom}, yields the
values of the condensates $\bsig_i(T_0,\mu_0)$ at the
potential minimum. For these values the condition ${\bf G}({\bf z}_0)=0$ with
${\bf z}_0 \left(\bsigx(T_0,\mu_0),\bsigy(T_0,\mu_0),T_0,\mu_0\right)$ is
obviously valid.
\item Prepare the derivative calculation for ${\bf H}({\bf z}_0)$ using\linebreak ADOL-C.
\item Evaluate the Taylor coefficients of $\bsig_i (T,\mu)$ at
$(T_{0},\mu_{0})$ up to the highest derivative degree $d$ desired by
the user.
\end{enumerate}
From now on, the Taylor expansions of $\bsig_i(T,\mu)$ around
$(T_0,\mu_0)$ are labeled as $\tsig_i^{(T_0,\mu_0,d)}(T,\mu)$. These
Taylor expansions are inserted in the grand potential which leads to
the definition
\begin{equation}
\tOmega^{(T_0,\mu_0,d)}(T,\mu) = \Omega\left(T,\mu;
  \tsigx^{(T_0,\mu_0,d)}(T,\mu), \tsigy^{(T_0,\mu_0,d)}(T,\mu) \right)
\,. 
\end{equation}

The function $\tOmega$ is exact in the explicit $\mu$-dependence but
only exact up to order $d$ in the implicit dependence. Thus, the
$k$-order derivatives of $\tOmega$ correspond to the derivatives of
the original $\bOmega$ if the derivatives are evaluated at the
expansion point $(T_0,\mu_0)$ and $k \leq d$, i.e., we have
\begin{equation}
\frac{d^k}{d\mu^k} \tOmega^{(T_0,\mu_0,d)}(T_0,\mu_0) =
\frac{d^k}{d\mu^k} \bOmega(T_0,\mu_0) \quad \text{for } k \leq d\ . 
\end{equation}
This equation is valid only at the point $(T_0,\mu_0)$. In order to
obtain the desired $\bOmega$ derivatives at another $(T,\mu)$ point,
the expansion coefficients of $\tsig_{q,s}$ have to be recalculated
for each $(T,\mu)$ point.

However, this reduces the problem of calculating the derivatives of
$\bOmega$ with only implicitly known functions $\bsig_{i}$ to the
calculation of the $\tOmega$ derivatives with explicitly known
$\tsig_{i}$.

Finally, the calculation of the derivatives of $\tOmega$ then requires
two steps:
\begin{enumerate}
\item Prepare the derivative calculation for $\tilde{\Omega}$ with \adolc at
$(T_0,\mu_0)$ which were chosen for the evaluation of the coefficients
of $\tsig_{i}$.
\item Evaluate the Taylor coefficients of $\tilde{\Omega}$.
\end{enumerate}

\subsection{Divided Differences}

Another method to approximate derivatives of a function is based on
divided differences which is explained in the following. 

Based on the definition of the derivative of a function $f$ at a point
$x$
\begin{equation}
\label{eq:def_deriv}
f'(x) = \lim_{h\rightarrow 0}\frac{f(x+h)-f(x)}{h} 
\end{equation}
the simplest linear approximation for $f'(x)$ is obtained by
calculating the right-hand side of \Eq{eq:def_deriv} for a small but finite
value of $h$
\begin{equation}
\label{eq:dd}
f'(x) \approx \frac{f(x+h)-f(x)}{h}\ .
\end{equation}

The problem with this approximation is that it involves two types of
errors (cf.~e.g.~\cite{NR}). If $h$ is too large, the so-called
truncation error induced by the used approximation or algorithm to
calculate the derivative becomes significant. 
On the other side, when $h$ becomes too small another error, 
the rounding error yields cancellations in the enumerator of (\ref{eq:dd}) 
and spoils the quality of the approximation.

Since the
two error sources compete with each other, one has to find an optimal
value of $h$ for which the numerical error of the derivative
evaluation is smallest. In general, this optimal $h$ varies with $x$,
the point at which the derivative is calculated.

The truncation error is relatively easy to control. By comparing
\Eq{eq:dd} with a Taylor expansion for $f(x+h)$ around $x$ one sees
that the truncation error of the linear approximation is of
${\cal O} (h)$, i.e., the error is a linear function of $h$. Thus,
decreasing $h$ will also decrease the truncation error. In addition,
by increasing the degree of the expansion the truncation error can
also be further improved. One such improved extrapolation, the
Richardson expansion, is based on the $n$-th order Taylor expansion
for $f(x+h)$ and $f(x-h)$ around $x$ for which a truncation error of
the order ${\cal O} (h^{2n})$ can be derived. By repeating the
algorithm for the determination of the truncation error, a better approximation for the first derivative $f'(x)$ can be obtained.
Similar improvements of the truncation error for higher-order
derivatives are also known.

\begin{figure*}[t!]
\centering
\subfigure[Second-order derivative]{
  \includegraphics[width=0.4\linewidth]{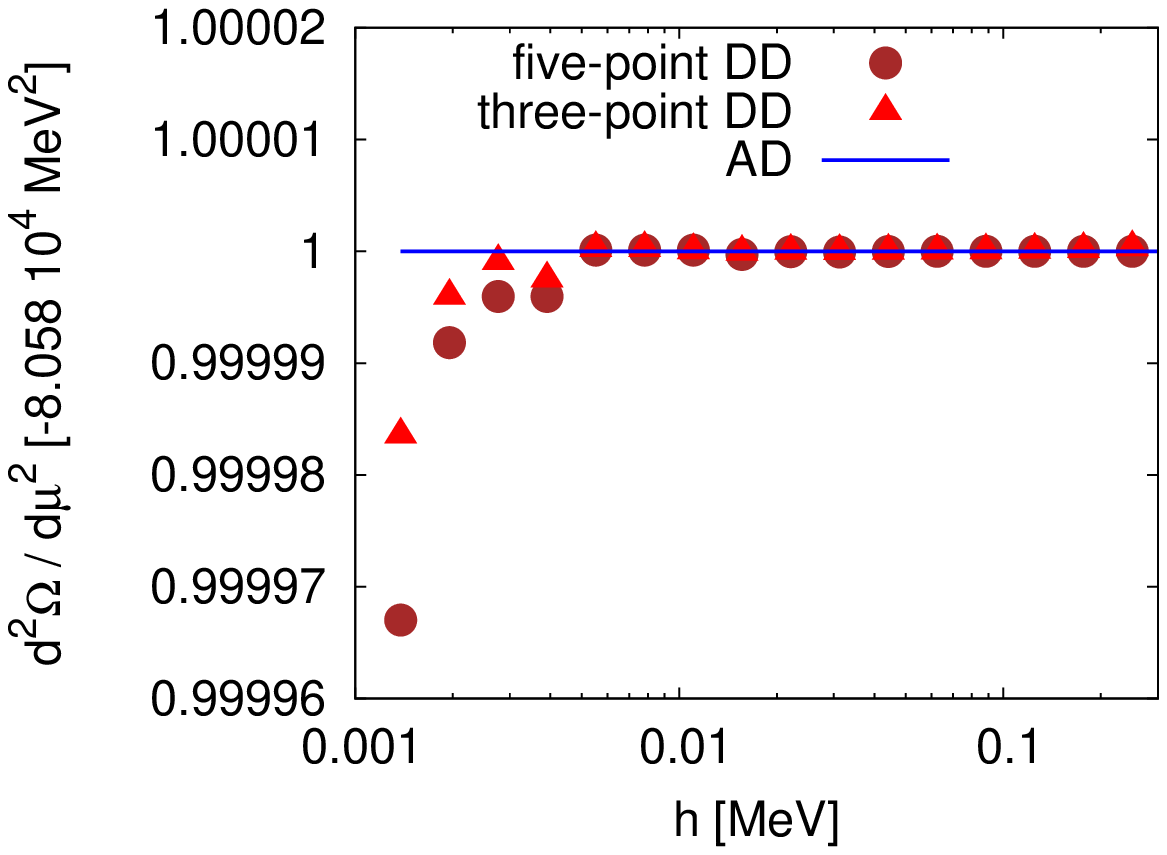}}
  \hspace{.1\linewidth}
\subfigure[Fourth-order derivative]{
  \includegraphics[width=0.4\linewidth]{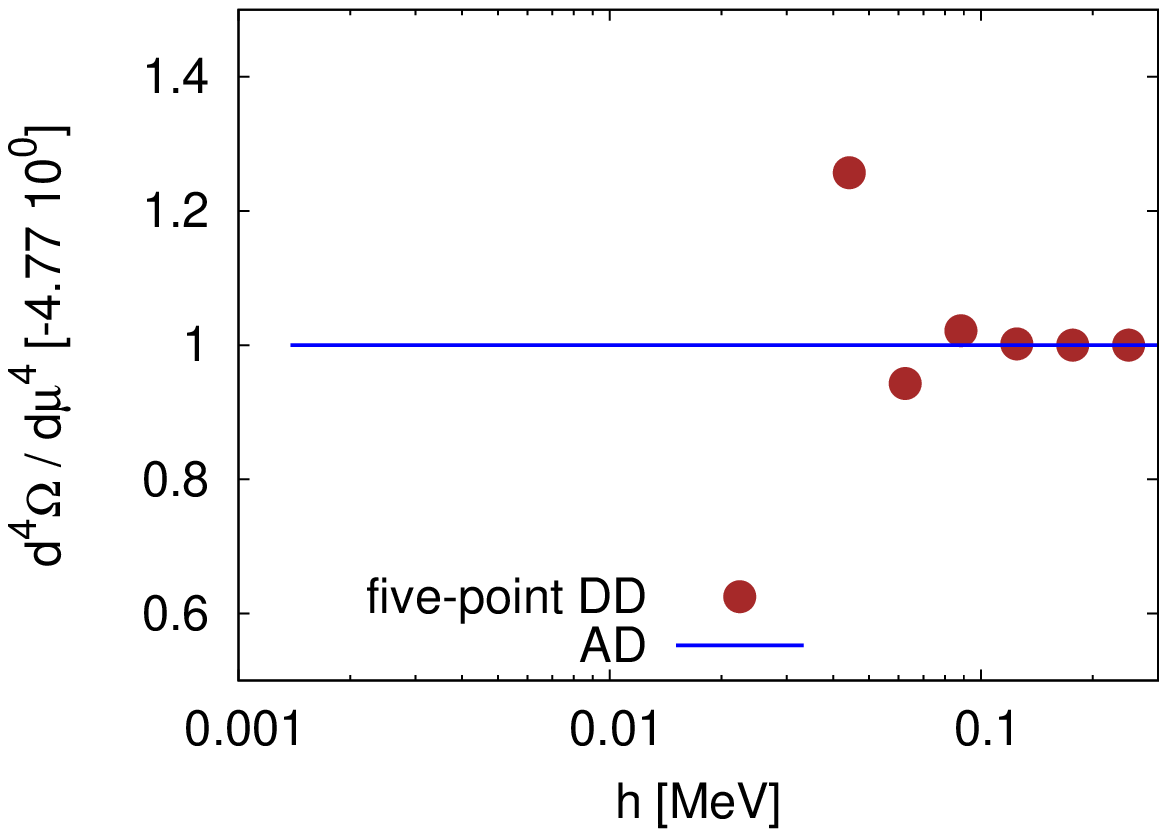}}
\caption{Comparison of the second- (left) and fourth-order (right)
  derivatives of the grand potential $\bOmega$, evaluated at
  $(T,\mu)=(183,0)$ MeV, with respect to $\mu$ using three points
  and five points DD with the AD technique. All DD results are
  normalized to the AD results.}
\label{fig:advsnum_t0}
\subfigure[Second-order derivative]{
\centering
  \includegraphics[width=0.4\linewidth]{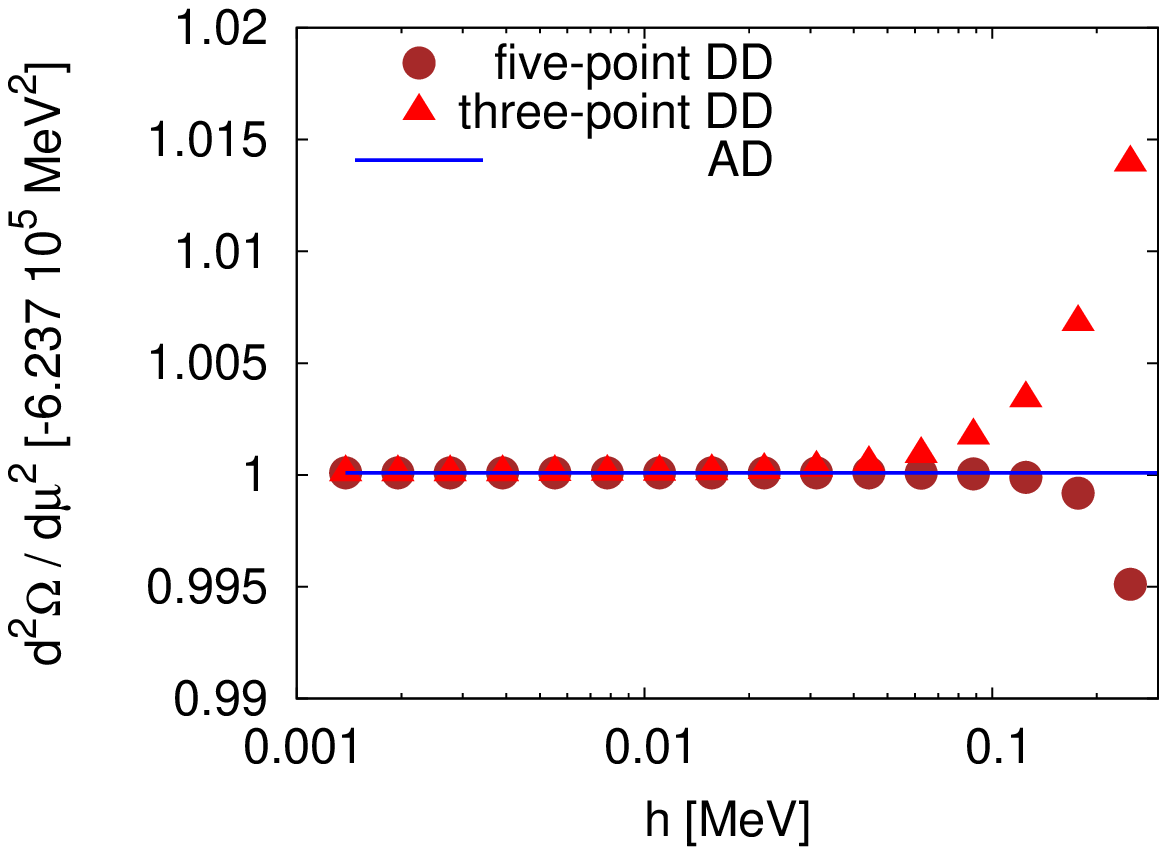}}
  \hspace{.1\linewidth}
\subfigure[Fourth-order derivative]{
  \includegraphics[width=0.4\linewidth]{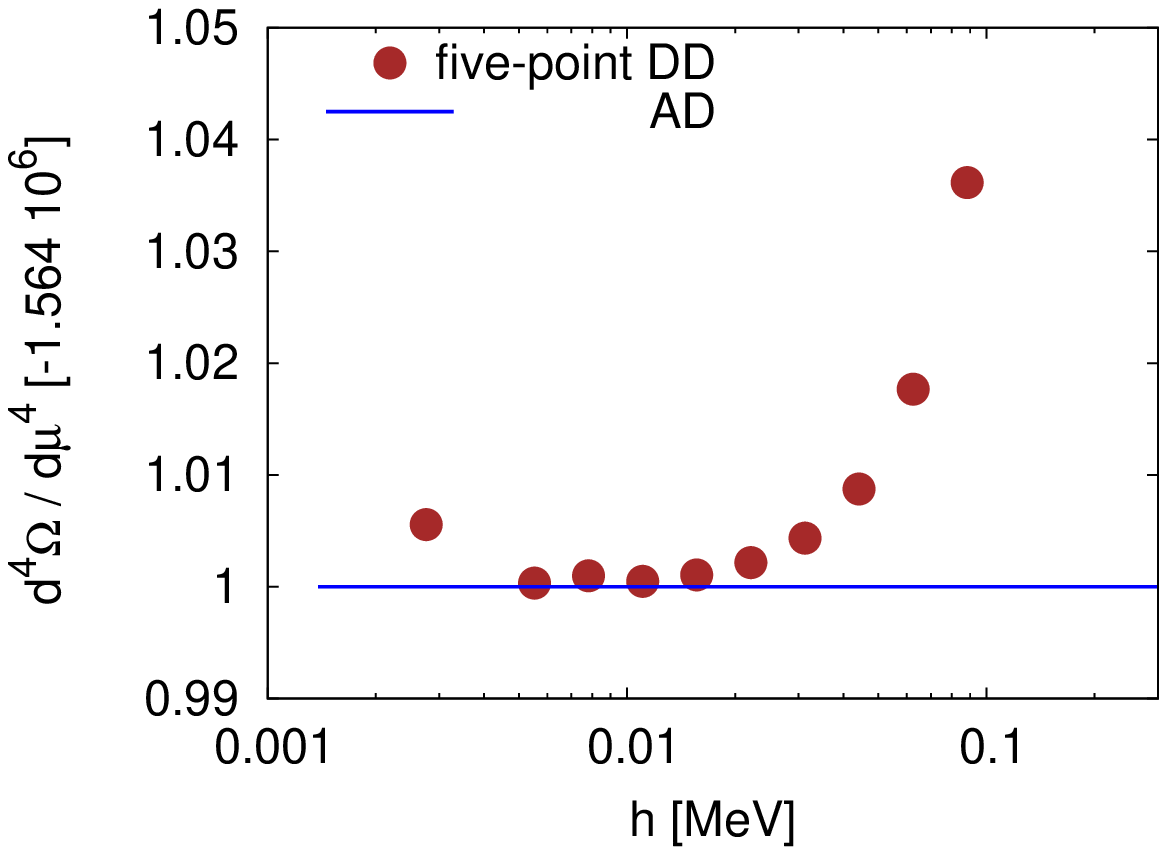}}
\caption{Similar to Fig.~\ref{fig:advsnum_t0} but $\bOmega$
  evaluated at another point in the phase diagram,
  $(T,\mu) =(63,327)$ MeV.}
\label{fig:advsnum_cep}
\end{figure*}

As an example the corresponding approximations for the second
derivative $f''(x)$ with three grid points $x_i$
\begin{equation}
\label{eq:2nd_three}
f''(x) = \frac{1}{h^2} \left[f(x_1) - 2 f(x_2) + f(x_3) \right] +
\mathcal{O}(h^{2})
\end{equation}
and with five grid points
\begin{multline}
\label{eq:2nd_five}
f''(x) = \frac{1}{12h^{2}}\left[ - f(x_{0}) +16 f(x_{1})\right.\\
\left. - 30
  f(x_{2}) + 16 f(x_{3})- f(x_{4})\right] + \mathcal{O}(h^{4}) 
\end{multline} 
are itemized. The grid points are given by
\begin{gather*}
x_{0} = x - 2 h, \qquad x_{1} = x -h, \quad x_{2} = x\\  x_{3} = x+h,
\quad x_{4} = x+2 h\ .
\end{gather*}
For completeness the fourth-order derivative is quoted
\begin{multline}
f''''(x)  = \frac{1}{h^4} \left[f(x_0) - 4 f(x_1) + 6 f(x_2) \right.\\ 
\left.- 4
f(x_3) + f(x_4) \right] +\mathcal{O}(h^{4})\ .
\end{multline}
where at least five grid points are needed for its calculation.

The disadvantage of such type of improvements is that the function has
to be evaluated at several different grid points $x_i$ which are
located in the vicinity of $x$.

The other error source, the rounding error, depends on the used format
of the floating point number representation in the computer. 
A single precision IEEE floating point number is stored in a 32-bit word, 
where 8 bits are used for the biased exponent and the fractional part of the
normalized mantissa is a 23-bits binary number. 
One bit in the IEEE format is always reserved for the sign of the number. 
A double precision number occupies 64 bits, with the biased exponent stored 
in 11 bits and the fractional part is stored on the remaining 52 bits.
Thus, besides the fact that one can represent only a finite subset of
all real numbers, all floating point calculations are
furthermore rounded resulting in incorrect values. 
The smallest positive number $\epsilon$, where the floating point approximation
for $1+\epsilon$ is indeed larger than one is called the machine
precision. 
When one rounds to the nearest representable number the machine precision
is roughly $\epsilon \sim 2^{-m}$ where $m$ is the number of bits used
to store the mantissa's fraction. 
For a single precision representation one finds $\epsilon \sim 2^{-23} \sim 10^{-7}$ 
and for a double precision number calculation
$\epsilon \sim 2^{-52} \sim 10^{-16}$. 
This means that single precision numbers have at most about 7 accurate digits 
while double precision numbers have about 16 accurate digits. 
But in general, due to the error propagation during the application of 
approximate algorithms the number of accurate digits for a numerical solution
decreases. 
Therefore, the rounding error will be several orders of
magnitude larger for a more complicated calculation such as the one
for the thermodynamic potential. 
To minimize this source of error in the derivative calculation of 
the thermodynamic potential, a larger value of $h$ is reasonable.

In order to estimate these numerical errors and verify the quality of
the \adolc evaluations the results of the derivative calculation
obtained with AD are confronted with the DD method.

In Figs.~\ref{fig:advsnum_t0} and \ref{fig:advsnum_cep} the results of
a DD evaluation as a function of $h$ in comparison with the AD
calculation for the second-order and fourth-order derivative of the
thermodynamic potential are shown. Fig.~\ref{fig:advsnum_t0} shows the
$\mu$-derivatives of the potential evaluated at $(T,\mu)=(183,0)$ MeV
which is close to the crossover phase transition in the $(T,\mu)$
phase diagram. One can clearly see the competition of the truncation
and rounding errors. For the second-order derivative the optimal
value is around $h \sim 0.05 \MeV$ while for the fourth-order derivative a
slightly larger value $h \sim 0.1 \MeV$ leads to more stable results. In
Fig.~\ref{fig:advsnum_cep} the same derivatives are calculated at the
point $(T,\mu)=(63,327)$ MeV which is near the critical end point in
the phase diagram. While in the previous Fig.~\ref{fig:advsnum_t0} the
rounding error dominates, the truncation error is now more important.
For the second-order derivative almost no rounding error is visible in
the resolution shown. Since the truncation error for the five-point
expression, \Eq{eq:2nd_five}, is of the order
$\mathcal{O} \left(h^4\right)$ and of the order
$\mathcal{O} \left(h^2\right)$ for the corresponding three-point
equation, \Eq{eq:2nd_three}, the results of the five-point derivative
is indistinguishable already for $h \sim 0.1 \MeV$ while for the
three-point formula a smaller value of $h \sim 0.02 \MeV$ is required. For
the fourth-order derivative the interval where the derivative does not
vary with $h$ is very small. Only for $h \sim 0.01 \MeV$ the DD result is
close to the AD result.

\begin{figure}[hbt!]
\begin{center}
\includegraphics[width=0.8\linewidth]{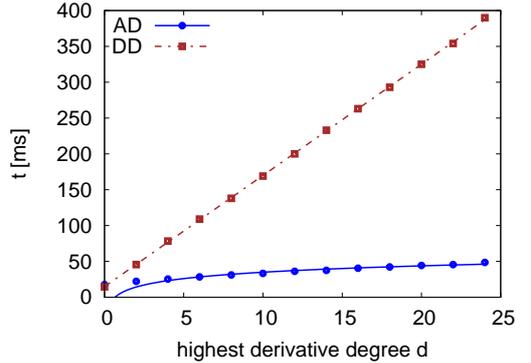}
\caption{Runtime comparison for the DD and AD approaches.}
\label{fig:mwad_runtime}
\end{center}
\end{figure}

In summary, one realizes that the DD derivatives require a very
careful fine-tuning of the $h$ value. The DD result coincides always with
the DD results where the $h$ variation vanishes. One finds that the AD
technique is more efficient than the DD method. The DD calculation
always requires the evaluation of the function at several points,
e.g., for the fourth derivative five function evaluations are
necessary. In our case this involves the solution of the
EoM at these five nodes. This is a time-consuming disadvantage of the
DD method. With the AD the EoM need to be solved only once. 
Despite the fact that it is required to generate an internal function representation 
of the evaluation of the EoM solution and of
the thermodynamic potential inside of \adolc, the AD implementation is much faster.
Corresponding runtime measurements are illustrated in Fig. \ref{fig:mwad_runtime}.

The runtime of the DD approach can be described by the linear function $f(d)=m*d+a$
where as the AD runtime performs like $g(d)=c*\log_2 (d)+b$.
This result fits perfectly to the computational complexity of the AD approach described in Sec.~\ref{sub:4.3}.

\section{Taylor coefficients}
\label{sec:taylor}

\begin{figure*}[t]
\centering
\includegraphics[width=.85\linewidth]{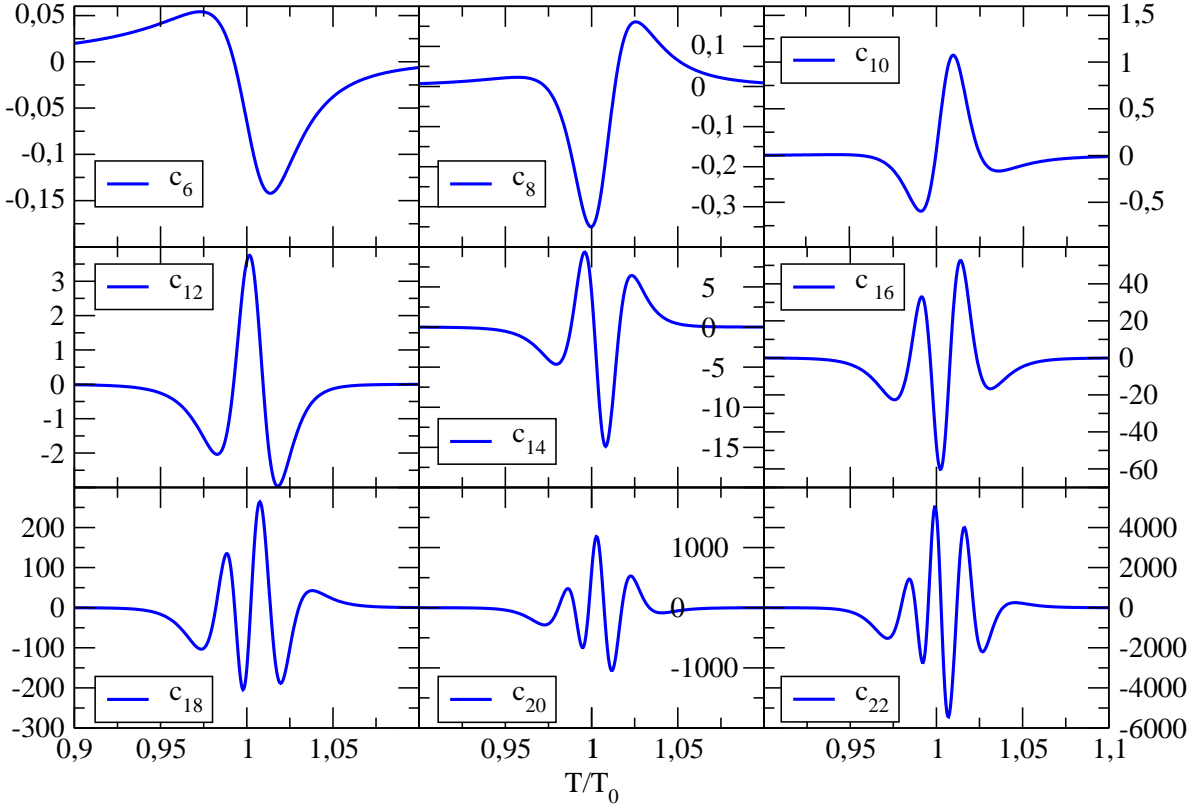}
\caption{Taylor expansion coefficients $c_{6}(T)$ to $c_{22}(T)$ as a
 function of the temperature (see text for details).}
\label{fig:adtaylor}
\end{figure*}

As previously illustrated, both error sources for a derivative
calculation with the DD method are in general difficult to keep under
control, in particular, if higher-order derivatives are involved.
However, with the AD method it is possible to obtain higher-order
derivatives with  very high precision. In the following an explicit
example is given within the already introduced linear sigma model.

Higher derivatives are required if one is interested, e.g., in the
extrapolation of Monte Carlo lattice simulations of strongly
interacting matter (lattice gauge theory) to finite quark chemical
potential. At finite quark chemical potential such types of Monte
Carlo simulations cannot be directly
performed~\cite{Philipsen:2007rj}. One possible extrapolation to
finite quark chemical potential is based on a Taylor expansion around
zero chemical potential~\cite{Allton:2002zi, Allton2005gk}.

For this purpose, we consider the same kind of expansion in the
quark-meson system described by the \lsm. An example is given by the
coefficients in the expansion of the pressure $p$ which is related to
the thermodynamic potential via
$p(T,\mu) = - \bOmega\left(T,\mu \right).$ At fixed temperature and
small values of the quark chemical potential the pressure may be
expanded in a Taylor series around $\mu=0$,
\begin{equation}
\frac{p(T,\mu)}{T^4} = \sum_{n=0}^\infty c_n(T)
\left(\frac{\mu}{T}\right)^n \ ,
\label{eq:pressuretaylor}
\end{equation}
where the expansion coefficients are given in terms of derivatives of
the pressure
\begin{equation}
c_n(T) = \left.\frac{1}{n!} \frac{\partial^n\left(p(T,\mu)
      /T^4 \right)}{\partial \left(\mu/T\right)^n}
\right|_{\mu=0}\ . 
\label{eq:taylorcoeffs}
\end{equation}
The series is even in $(\mu/T)$ which reflects the invariance of
the partition function under the exchange of particles and
antiparticles.

In Fig.~\ref{fig:adtaylor} the expansion coefficients $c_6(T)$ to
$c_{22}(T)$ are shown as function of the scaled temperature $T/T_0$.
Here, $T_0$ is the pseudocritical temperature at which the crossover
transition occurs for vanishing chemical potential. Since the first
three expansion coefficients $c_0, c_2$ and $c_4$ are already known and
well-understood we do not show them again~\cite{Philipsen:2007rj}. In
lattice gauge theory one can currently calculate the first five
coefficients, $c_0, \ldots, c_8$~\cite{Miao:2008sz}.

The higher coefficients $c_{n}$ with $n>4$ vanish for temperatures
basically outside of a five percent window around $T_0$. Thus, all
coefficients are only shown in the range $0.9 <T/T_0 <1.1$. All curves
are smooth oscillating functions around zero even up to the
$22^\text{nd}$ derivative order. The amplitude of the oscillation and
the number of roots around $T_0$ increases with the order $n$. Thus,
this oscillating behavior of the coefficients obviously requires a
smaller $h$ in order to decrease the truncation error but then the
rounding error increases. Already in this example the error sources
are dramatic for such a high degree of derivatives. Therefore it is
not reasonable and actually not possible to obtain the higher coefficients
with standard techniques such as the DD method.

\section{Summary}
\label{sec:summary}

A novel numerical technique, which is based on algorithmic
differentiation, for the calculation of arbitrarily high-order and
high-precision derivatives has been presented. The new feature of the
technique is the additional treatment of implicitly defined functions.
In addition, the basic concepts of the algorithmic differentiation for
explicit dependencies is discussed.

As a demonstration of the successful extension to implicitly defined
functions the AD technique is applied to a quantum-field theoretical
model for strongly-interacting matter. In this model the implicitly
defined functions are represented by the underlying equations of
motion where the implicitly defined order parameter is known only
numerically.

Two important error sources namely, the rounding and truncation error, for a
derivative calculation in general are discussed in detail.
Furthermore, the results with the improved AD method are confronted to
those obtained by standard divided difference (DD) methods. In the
comparison the rounding and truncation errors can clearly be
identified. While for a second-order derivative calculation the
error sources are still controllable, they become intractable for
higher orders.

In the model example higher-order derivative coefficients of a Taylor
expansion for the pressure are calculated up to $22^\text{nd}$ order.
Since these coefficients are calculated for the first time, no
comparison with other results can be performed. The obtained curves
are very stable and smooth functions which demonstrates the power of
the novel AD technique. In a forthcoming publication
\cite{Taylorcoeff} this method will be applied to the more realistic
Polyakov-Quark-Meson model for three quark flavors
\cite{Schaefer:2008ax,Schaefer:2009ui}.

The presented AD technique augmented by implicitly defined
dependencies can be applied to a wide class of problems, where
high-order derivatives are involved. Standard alternative methods for
the derivative calculation such as the DD method fail due to
uncontrollably increasing errors. Especially, in the case of only
numerically known implicit dependencies, an analytic solution is actually
not possible. Here, the AD method is still applicable and displays its
exceptional impact.

\paragraph{Acknowledgment}
The work of MW was supported by the Alliance Program of the Helmholtz
Association (HA216/ EMMI) and BMBF grants 06DA123 and
06DA9047I. We thank J.~Albersmeyer, F.~Karsch,
A.~Krassnigg, R.~Roth and J.~Wambach for useful discussions and
comments.

\end{document}